\documentclass[twocolumn,aps,prstab]{revtex4-1}

\usepackage{etoolbox,graphicx,framed,amssymb,mathrsfs,hyperref}
\usepackage{lineno,textcomp,fancyhdr,calligra,ifsym,amsmath,amsthm,mathtools,amsmath}

\begin{document}

\title{{\bfseries\large Thomas-BMT equation generalized \\ to electric dipole moments and field gradients}}

\author{Eric M. Metodiev}
\affiliation{Harvard College, Harvard University, Cambridge, MA 02138, USA
\\Center for Axion and Precision Physics Research, IBS, Daejeon 305-701, Republic of Korea
\\Department of Physics, KAIST, Daejeon 305-701, Republic of Korea}
\renewcommand{\[}{\begin{equation}}
\renewcommand{\]}{\end{equation}}
\renewcommand{\v}{{\bf v}}
\newcommand{\E}{{\bf E}}
\newcommand{\B}{{\bf B}}
\newcommand{\s}{{\bf s}}
\newcommand{\M}{{\bf M}}
\newcommand{\N}{{\bf N}}
\newcommand{\R}{{\bf R}}
\renewcommand{\b}{\boldsymbol\beta}
\renewcommand{\d}{{\bf d}}
\newcommand{\w}{{\bf w}}
\newcommand{\m}{\boldsymbol\mu}

\begin{abstract}
An expression is presented for the relativistic equations of motion, including field gradients, of a particle and its spin with electric and magnetic dipole moments aligned along the spin axis. An electromagnetic duality transformation is used to generalize a Thomas-BMT equation with gradient terms. Corrections to particle dynamics in storage rings for precision $(g-2)$ and electric dipole moment measurements are calculated, and applications to precision particle tracking programs are considered.
\end{abstract}
\maketitle
\section{Introduction}

A detailed knowledge of the spin dynamics of particles with non-zero electric dipole moments (EDMs) and magnetic dipole moments (MDMs) is necessary for precision EDM and $(g-2)$ measurements using spin precession in storage rings \citep{pedm,sredm4,sredm1,sredm2,srmdm3}. The Thomas-Bargmann-Michel-Telegdi (T-BMT) equation \citep{thomas,tbmt} governs the classical spin dynamics of a particle with a non-zero MDM in electric and magnetic fields, neglecting field gradients. Recently, derivations have been presented which generalize the T-BMT equation to include a non-zero particle EDM based on duality transformations \citep{dual1,spinpre} and explicit relativistic constructions \citep{spinpre2}.

The spin equation of motion of a particle with a non-zero MDM including first order field gradients has also been established \citep{good}. By making use of an electromagnetic duality transformation on these equations of motion, a generalization of the T-BMT equation for non-zero particle EDMs and MDMs and first order field gradients is determined. The corrections to the spin and particle equtions of motion are then studied. We find that typical experimental methods in storage ring EDM methods are robust to higher order than previously demonstrated. Higher-order corrections to the dynamical equations used in many precision particle tracking programs \citep{tracking2,tracking3} are also presented.

\section{Dynamics with a non-zero magnetic dipole moment}
R.H. Good \citep{good} determined the classical equations of motion for a particle and its spin with a MDM $\mu$ along its spin direction to first order in the field gradients. For a particle of mass $m$, electric charge $e$, spin angular momentum $I \hbar$, and velocity $\v = c\b$, he derived that:

\[\begin{aligned}\label{eq1}mc\frac{d(\gamma \b)}{dt}&= e \E + ec \b\times\B  \\& + \frac{\mu\gamma}{cI\hbar}[\nabla + \b\times(\b\times\nabla)+ \frac{1}{c}\b\partial_t]\s\cdot\R
\end{aligned}\]
and
\[\begin{aligned}
\frac{d\s}{dt} &= \frac{\mu}{cI\hbar}\frac{1}{\gamma}\s\times\R - \frac{e}{mc}\s\times\N
\\& + \frac{\mu}{I\hbar mc^2}\frac{1}{\gamma + 1} \s \times (\b\times \nabla)[\s \cdot \R] ,
\end{aligned}\]
where:
\[\begin{aligned}
&\M =c \B - \frac{\gamma}{\gamma + 1} \b \times \E\\
&\N = \frac{\gamma}{\gamma + 1}(\E +c\b \times\B)\times \b\\
&\R = \M + \gamma \N.
\end{aligned}\]

A non-zero quadrupole moment $q$ is also considered in the solution, but we omit it here for clarity. This method can also be used to extend the T-BMT equation to electric quadrupole moments. We define the EDM $\d$ and MDM $\m$ in terms of the rest frame spin $\s$ as:
\[\d = \frac{\eta}{2}\frac{e}{mc}\s,\,\,\,\,\,\,\m = \frac{g}{2}\frac{e}{m}\s,\]
where these relations define $g$ and $\eta$.

We can also write the Equation \ref{eq1} as an equation for $\b$ in the form:
\[\begin{aligned}\gamma mc\frac{d\b}{dt}&= e \E + e c\b\times\B - e \b(\b \cdot\E)  \\& + \frac{\mu\gamma}{cI\hbar}[\nabla + \b\times(\b\times\nabla)+ \frac{1}{c}\b\partial_t]\s\cdot\R.
\end{aligned}\]

To write the equations in covariant form, we use the four-velocity $u^\mu$ and define the spin 4-pseudovector which takes the form $a^\mu = (0,\s)$ in the particle rest frame. Further we use the electromagnetic field strength tensor $F^{\mu\nu}$ and its dual $F^{*\mu\nu}$. In these terms, the equations become:
\[m\frac{du_\nu}{d\tau} = e F^{\nu\rho}u_\rho + \frac{\mu}{I\hbar c} u_\sigma \left( \partial_\nu +\frac{1}{c^2} u_\nu u^\gamma  \partial_\gamma \right)F^{* \sigma \rho} a_\rho\]

\[\begin{aligned}\frac{d a_\nu}{d\tau} &= \frac{\mu}{I\hbar}F_{\nu \sigma}a^\sigma - \frac{1}{c^2}\left(\frac{\mu}{I\hbar} - \frac{e}{m}\right) u_\nu u_\mu F^{\mu\sigma}a_\sigma \\& + \frac{\mu}{I\hbar mc^3} u_\nu u_\mu a^\gamma \partial_\gamma F^{*\mu\sigma}a_\sigma.\end{aligned}\]

It can be checked explicitly that when the gradient terms are neglected, the equation for $\dot\s$ reduces to the T-BMT equation and the equation for $\dot\b$ reduces to the Lorentz force law for a point particle \citep{jackson}.

The higher-order T-BMT equation is a classical, relativistic equation, which omits quantum corrections to the dynamics \citep{good}. Quantum corrections due to the external fields can be the same order of magnitude as the classical corrections due to non-zero field gradients. These corrections must thus be dealt with in addition to the gradient terms discussed here. This extension can be accomplished by applying the duality procedure to more fully quantum treatments \citep{jacek}. We will neglect such terms in our discussion and focus our attention to generalizing the classical equations and determining their impact on storage ring dynamics.

\section{Duality Transformation}
The Maxwell equations with charge sources are invariant under the following electromagnetic duality transformation \citep{dual,dual2}:
\[\begin{aligned}
&\E \rightarrow - c \B
&\B\rightarrow \E/c\\
&\rho_e \rightarrow - \rho_m/c
&\rho_m \rightarrow c\rho_e,
\end{aligned}\]
as well as the general rotational duality transformation. Similarly, in covariant form the duality transformation becomes $F_{\mu\nu}\rightarrow -F^*_{\mu\nu}$ and $F^{*}_{\mu\nu}\rightarrow  F_{\mu\nu}$.

Suppose then that we have the equations of motion for the particle and spin dynamics for a particle of magnetic dipole moment $\m$ in an electric field $\E$ and magnetic field $\B$. The transformation taking $\m \rightarrow c \d$, $\E \rightarrow -c\B$, and $\B\rightarrow \E/c$ will then yield the equations of motion for a particle of electric dipole moment $\d$ in electromagnetic fields. Alternatively, we can map $eg\rightarrow e\eta$ rather than $\m \rightarrow c\d$.

Note that the spin equation expresses the spin $\s$ in the rest frame while the fields $\E$ and $\B$ are those in the laboratory frame. The duality transformation repects the Lorentz transformation of the electromagnetic field:
\[\begin{aligned}&{\E'}_\parallel = \E_\parallel  
&{\E'}_\perp = \gamma(\E_\perp + c\b \times \B)\\& {\B'}_\parallel = \B_\parallel & {\B'}_\perp = \gamma(\B_\perp - \frac1c\b\times\E), \end{aligned}\]
for a transformation from the unprimed lab to the primed rest frame. The results derived for the rest frame and lab frame spin dynamics of a particle with non-zero EDM and MDM must be invariant under the duality transformation.

\section{Dynamics with non-zero electric and magnetic dipole moments}
We now extend the equations of motion with only an MDM to a set of equations which are invariant under the duality transformation and which reduce properly in the $\d = \m = 0$ limit. Further we assume that there are no terms which depend on both $g$ and $\eta$. For simplicity, we take the particle to have zero magnetic charge, though this assumption can be easily relaxed. In covariant form, we then have:

\[\begin{aligned}m\frac{du_\nu}{d\tau} &= e F^{\nu\rho}u_\rho\\& + \frac{\mu}{I\hbar c} u_\sigma \left( \partial_\nu +\frac{1}{c^2} u_\nu u^\gamma  \partial_\gamma \right)F^{* \sigma \rho} a_\rho
\\&+ \frac{d}{I\hbar} u_\sigma \left( \partial_\nu +\frac{1}{c^2} u_\nu u^\gamma  \partial_\gamma \right)F^{\sigma \rho} a_\rho\end{aligned}\]

\[\begin{aligned}\frac{d a_\nu}{d\tau} &= \frac{\mu}{I\hbar}F_{\nu \sigma}a^\sigma - \frac{1}{c^2}\left(\frac{\mu}{I\hbar} - \frac{e}{m}\right) u_\nu u_\mu F^{\mu\sigma}a_\sigma \\& + \frac{\mu}{I\hbar mc^3} u_\nu u_\mu a^\gamma \partial_\gamma F^{*\mu\sigma}a_\sigma
\\&- \frac{c d}{I\hbar}F^*_{\nu \sigma}a^\sigma + \frac{d}{I\hbar c}  u_\nu u_\mu F^{*\mu\sigma}a_\sigma \\& + \frac{d}{I\hbar mc^2} u_\nu u_\mu a^\gamma \partial_\gamma F^{\mu\sigma}a_\sigma.\end{aligned}\]

From this we find that the particle velocity and its spin must obey the following equations:
\[\begin{aligned}\label{big1}mc\frac{d(\gamma \b)}{dt}&= e \E + ec\b\times\B  \\& + \frac{\mu\gamma}{cI\hbar}[\nabla + \b\times(\b\times\nabla)+ \frac{1}{c}\b\partial_t]\s\cdot\R  \\& + \frac{d\gamma}{I\hbar}[\nabla + \b\times(\b\times\nabla)+ \frac{1}{c}\b\partial_t]\s\cdot\tilde \R
\end{aligned}\]
\[\begin{aligned}\label{big2}
\frac{d\s}{dt} &= - \frac{e}{mc}\s\times\N
\\&+ \frac{\mu}{cI\hbar}\frac{1}{\gamma}\s\times\R + \frac{\mu}{I\hbar mc^2}\frac{1}{\gamma + 1} \s \times (\b\times \nabla)[\s \cdot \R]\\& + \frac{d}{I\hbar}\frac{1}{\gamma}\s\times\tilde \R
 + \frac{d}{I\hbar mc}\frac{1}{\gamma + 1} \s \times (\b\times \nabla)[\s \cdot\tilde\R],
\end{aligned}\]
where we have defined the transformed quantities:
\[\begin{aligned}
&\tilde \M =\E + c \frac{\gamma}{\gamma + 1} \b \times \B\\
&\tilde\N = \frac{\gamma}{\gamma + 1}(-c\B + \b \times\E)\times \b\\
&\tilde \R = \tilde \M + \gamma\tilde\N.
\end{aligned}\]

Similarly, we can rewrite Equation \ref{big1} as:
\[\begin{aligned}\gamma mc\frac{d\b}{dt}= &e \E + ec \b\times\B - e\b(\b\cdot\E)  \\& + \frac{\mu\gamma}{cI\hbar}[\nabla + \b\times(\b\times\nabla)+ \frac{1}{c}\b\partial_t]\s\cdot\R  \\& + \frac{d\gamma}{I\hbar}[\nabla + \b\times(\b\times\nabla)+ \frac{1}{c}\b\partial_t]\s\cdot\tilde\R.
\end{aligned}\label{corrpar}\]

Note that the quantities $\R$ and $\tilde\R$ can be expressed in terms of the fields as:
\[\begin{aligned}
&\R =  \gamma c \B - \gamma \b \times \E - c \frac{\gamma^2}{\gamma + 1}(\b \cdot\B)\b = c\B'\\
&\tilde \R =  \gamma \E + \gamma c \b \times\B - \frac{\gamma^2}{\gamma + 1}(\b\cdot \E)\b = \E',
\end{aligned}\]
which we identify as the Lorentz transformed $\B$ and $\E$ fields measured in the particle rest frame, as expected.

Considering the case of a spin-$\frac12$ particle where $I = \frac12$ and neglecting the field gradients contributions, we find that the equation of motion for the spin reduces to:
\[\begin{aligned}\frac{d\s}{dt}& = \frac{e}{m}\s\times \left[ \left(a + \frac{1}{\gamma}\right)\B \right. \\ & \left. - \frac{a\gamma}{\gamma + 1}\b(\b \cdot\B) -\left(a + \frac{1}{\gamma + 1}\right)\frac{\b\times\E}{c}\right. \\ & \left.+ \frac{\eta}{2}\left(\frac{\E}{c} + \b\times\B - \frac{\gamma}{\gamma + 1}\frac{\b(\b \cdot\E)}{c}\right)\right],\end{aligned}\]
where $a = (g-2)/2$. The generalized result thus properly reduces to the previously derived expression for the T-BMT equation with a non-zero EDM \citep{spinpre}.

\section{Storage Rings and Precision Particle Tracking}

There are several ongoing efforts to detect EDMs of charged particles with novel sensitivity by studying their spin precession in storage rings \citep{pedm,sredm4}. The governing dynamics of these precision measurements are the T-BMT equations, generalized to include EDMs. Additionally, measurements of the muon $(g-2)$ rely on the T-BMT equation \citep{srmdm3}. Understanding the gradient corrections to the T-BMT equation is important for present and future precision measurements.

The usual conditions for store ring EDM and $(g-2)$ experiments include the particle moving transversely to the electric and magnetic fields with both $\b\cdot\E = 0$ and $\b \cdot \B = 0$. A full treatment of these dynamics, without including field gradients, was performed by Fukuyama and Silenko \citep{spinpre}. We thus restrict our considerations to gradient terms contributing to $\dot\s$, which are:
\[\begin{aligned}\left(\frac{d\s}{dt}\right)_\nabla =&  \frac{\mu}{I\hbar mc^2}\frac{1}{\gamma + 1} \s \times (\b\times \nabla)[\s \cdot \R]\\&+ \frac{d}{I\hbar mc}\frac{1}{\gamma + 1} \s \times (\b\times \nabla)[\s \cdot\tilde\R].\end{aligned}\]

Conditions are further simplified for experiments using the frozen spin method \citep{sredm1,sredm2,frospin1,frospin2}, with the spin locked along the direction of the momentum. In this method, the particle energy, electric field, and magnetic field are chosen such that:
\[a\B - \left(a - \frac{1}{\gamma^2 - 1}\right)\frac{\b \times \E}{c} = 0,\]
cancelling the $(g-2)$ precession of the particle.

Neglecting the precession due to a non-zero $\eta$, the spin and momentum can be initially locked with $\s\times\b = 0$. This condition gives:
\[\begin{aligned}
&\s\cdot\R =  \gamma c \,\s\cdot\B- c \frac{\gamma^2}{\gamma + 1}(\b \cdot\B)(\s\cdot\b) = c\,\s\cdot\B\\
&\s\cdot\tilde \R =  \gamma\,\s\cdot \E - \frac{\gamma^2}{\gamma + 1}(\b\cdot \E)(\s\cdot\b) = \s\cdot \E,
\end{aligned}\]
which vanish with the spin along the momentum direction.

Since $\b\cdot \B = 0$ and $\b \cdot \E = 0$ for all points in the cross section of the ring, we find that:
\[\begin{aligned}&(\b\times\nabla)[\s\cdot\R] = 0, &(\b\times\nabla)[\s\cdot\tilde \R] = 0,\end{aligned}\]
which cancels the field gradient corrections to the spin dynamics. We see that the frozen spin method is robust to higher order corrections in the dynamics.

Thus storage ring experiments using the frozen spin method, with spin along the momentum direction, have vanishing corrections to first order in the classical field gradients. The quantum gradient corrections may similarly vanish, though they must be evaluated explicitly.


For the currently considered parameters of storage ring EDM experiments \citep{pedm2}, the effects due to gradient terms in $\dot\b$ and $\dot\s$ are negligible compared to the experimental sensitivities. We find that the frozen spin condition cancels the entire contribution of the gradient terms to the spin dynamics of the particles.

To relax the idealizations of the calculations, it is valuable to use precision particle tracking, which can evaluate the robustness of the experimental methods. For instance, there may be non-ideal fields with large gradients and $\boldsymbol\beta\cdot{\bf E}\neq 0$, such as realistic deflector electric fringe fields that have recently been implemented in tracking programs \citep{fringes}. Precision tracking can be used to evaluate the contributions of the higher order terms with the relaxed conditions including transverse particle momentum components and imperfectly frozen spins. The governing spin and particle dynamics with higher-order classical gradient and EDM terms are captured in Equations \ref{big2} and \ref{corrpar}.

\section{Conclusions}
Equations \ref{big1} and \ref{big2} express the classical relativistic equations of motion of a particle and its spin with a non-zero EDM and MDM in electric and magnetic fields $\E$ and $\B$ to first order in the field gradients. The spin equation of motion presents a further generalization of the T-BMT equation to include first order gradients and a non-zero EDM. By the same duality procedure, a non-zero magnetic charge of the particle can also be included. Using the full result for the generalized T-EDM \citep{good}, this framework can also accomodate electric and magnetic quadrupole terms.

These equations allow for more detailed analytical and precision tracking investigations into the spin and particle dynamics with non-zero EDMs. Contained in Equations \ref{big2} and \ref{corrpar}, the governing dynamics are developed for including the higher-order terms in the framework of presently existing precision particle tracking programs. The conditions of the frozen spin method, important to many precision EDM measurements in storage rings, were found to remain valid through first order in the classical gradient terms for both an EDM and MDM. The sizes of non-zero effects due to these terms are found to be below experimental sensitivities for the parameters of current storage ring EDM efforts. 

\newpage
\section*{Acknowledgements}
The author is grateful to Y.K. Semertzidis for his insightful comments. Special thanks are given to the Storage Ring EDM collaboration. DOE partially supported this project under BNL Contract No. DE-SC0012704. IBS-Korea partially supported this project under system code IBS-R017-D1-2014-a00.

\end{document}